# Electric Road Systems for Smart Cities: A Scalable Infrastructure Framework for Dynamic Wireless Charging

**Rishit Agnihotri*, Amit Chaurasia***

*Department of Computer and Communication Engineering, Manipal University Jaipur, India
Rishit.23fe10cce00051@muj.manipal.edu
Amit.chaurasia@jaipur.manipal.edu

## Abstract

*Dynamic Wireless Charging (DWC) systems using Electric Road Systems (ERS) can resolve problems associated with limited charging infrastructure and range anxiety that occur from electric vehicles being driven in a city environment. This paper presents the development of a scalable Modular Architecture for an ERS that consists of inductive power transfer, Vehicle-to-Infrastructure (V2I) communication, and artificial intelligence (AI)-based energy management. The performance evaluation of ERS was conducted using simulations in SUMO and MATLAB for an Indian urban corridor with very high densities. The evaluation indicated reductions of up to 30–35% in range anxiety, up to 40% in the number of deep discharge cycles, and improved stability of the grid (within the range of ±1.5%). The paper also provides a case analysis for Delhi, providing information on the economic feasibility of an ERS and establishing viable deployment strategies for ERS as a suitable solution for the future infrastructure of mobility in smart cities.*

## 1 Introduction

The experience of accelerated urbanisation, combined with the growing significance of climate change, has pushed cities across the globe to revamp their transport infrastructures. Electrification of mobility, particularly through electric vehicles (EVs), is amongst the key strategies towards decarbonising city transport. However, among the key bottlenecks toward the widespread adoption of EVs is the lack of existing static charging infrastructures, which tends to induce range anxiety, grid overload during peak hours of charging, and land use issues in dense cities [1].

Dynamic Wireless Charging (DWC) through Electric Road Systems (ERS) is an eco-friendly approach. DWC allows for EV charging on the move through inductively embedded coils placed under road surfaces, thus enabling instant energy transfer without the need to stop or direct user engagement. The technique of energy distribution reduces the dependence on gigantic, centralised charging nodes and encourages more equitable load distribution in urban energy networks. In addition, it aligns with the principles of smart cities through the unification of infrastructure with advanced systems.

This article proposes a modular and scalable architecture of an Electric Road System (ERS) that can be implemented in smart cities, particularly in developing countries like India, where traffic congestion, urbanisation, and energy infrastructure-related problems are prevalent. The system suggested integrates vehicle-to-infrastructure (V2I) communication, grid management with artificial intelligence, and renewable energy sources to formulate a smart and robust urban mobility solution. The foundation of the system design is the modularity, interoperability, and sustainability principles.

The objective of this research was to investigate the validity and performance potential of a viable, scalable Electric Road System (ERS) framework for providing dynamic wireless charging capability in a dense city environment.
The primary research objectives are as follows:

1. Can the ERS provide a solution to resolving range anxiety and the negative effects of battery degradation experienced by electric vehicles (EVs) when subjected to heavy traffic congestion?
2. What is the impact on the stability of the power grid due to the demand for energy from the ERS system?
3. What is the potential for creating a viable economic environment for deploying an ERS in India's smart cities?

Hypothesis: The hypothesis of the study is that an AI-enabled modular ERS architecture will provide a reduction in EV range anxiety of 30% or greater while maintaining stability of the power grid within a range of +/−2%.

## 2 Literature Survey

Several international projects have led the way for ERS deployment. Sweden's eRoad Arlanda project proved a 2 km electrified road with conductive rails set into the road, supplying up to 800 V DC to heavy trucks [2]. Israel's Electreon wirelessly supplies power to receivers in EVs with inductive coils set into the asphalt [3]. South Korea's OLEV (Online Electric Vehicle) project also implemented dynamic charging for buses and logistics fleets successfully [4].

Despite technological success, these pilots are faced with high infrastructure costs, poor interoperability, and regulatory uncertainty. Though research at institutions like KAIST and Stanford has addressed power transfer efficiency and coil design [5], deployment strategies at the city scale are not extensively researched. One of the principal gaps is integrating these systems into city-scale energy management, traffic systems, and policy frameworks.
Literature has addressed inductive charging physics [6], charging coil placement optimisation [7], and vehicle routing with DWC constraints [8]. Yet, few end-to-end frameworks consider urban-scale deployment, particularly in heterogeneous traffic in developing nations. This paper builds on the previous work by suggesting an end-to-end ERS system architecture with deployability in the real world, employing Delhi as a representative testbed.

| Study / Project | Technology Used | Scale of Deployment | Key Features | Limitations | Research Gap Addressed in This Work |
|---|---|---|---|---|---|
| eRoadArlanda (Sweden) | Conductive rail-based ERS | Pilot (2 km highway) | High power transfer (up to 800 V DC), suitable for heavy vehicles | Mechanical wear, safety concerns, limited interoperability | Lack of scalable, modular, and city-compatible architecture |
| Electreon (Israel) | Inductive Wireless Charging | Urban pilot roads | Wireless power transfer, minimal visual infrastructure | High installation cost, limited grid integration | Absence of AI-based grid optimisation and smart energy control |
| OLEV (South Korea) | Inductive charging for buses | Public transport system | Proven for buses, reduced battery size requirement | Limited to fixed routes, lacks adaptability to mixed traffic | No support for heterogeneous urban traffic environments |
| Choi et al. (2017) | Wireless Power Transfer (WPT) | Lab-scale / prototype | High efficiency coil design, strong theoretical foundation | Not validated at city scale | No deployment strategy or system integration |
| Kesler et al. (2014) | Dynamic WPT system | Prototype | Early implementation of dynamic charging | Low scalability, outdated technology assumptions | No AI integration or real-time optimisation |
| Zhang et al. (2023) | Speed optimisation for DWC | Simulation-based | Improved charging efficiency via speed control | Focus limited to vehicle dynamics only | Ignores infrastructure and grid-level impacts |
| Sun et al. (2023) | Coil misalignment analysis | Analytical study | Addresses efficiency loss due to misalignment | No system-level solution proposed | Does not integrate mitigation strategies in deployment |
| **This Work (Proposed ERS Framework)** | Inductive DWC + AI-based EMS + V2I + Smart Grid Integration | City-scale simulation (Indian urban corridor) | Modular architecture, AI-driven energy management, renewable integration, real-world case study (Delhi) | Simulation-based validation (no real-world prototype yet) | Provides end-to-end scalable, deployable ERS framework for smart cities |

## 3 System Architecture

Our ERS design is modular, scalable, and interoperable. It is composed of three main layers:

### 3.1 Physical Layer

The core of the system is the physical positioning of inductive coils on road segments. The coils, supplied by roadside power converters and substations, generate alternating electromagnetic fields that induce a current in the receiving coil of the EV. Modular segmental coils (2–5 m each) can be switched on separately depending on vehicle detection to minimise standby energy loss and electromagnetic interference (EMI) [9]. The embedded infrastructure should be weather-resistant and easy to maintain.

This layer also has integrated pavement sensors, protective shields, and thermal management systems. Coil operation depends on road temperature, vehicle clearance, and lateral position. To counteract this, this paper proposes dual-coil rows with adaptive phase control.

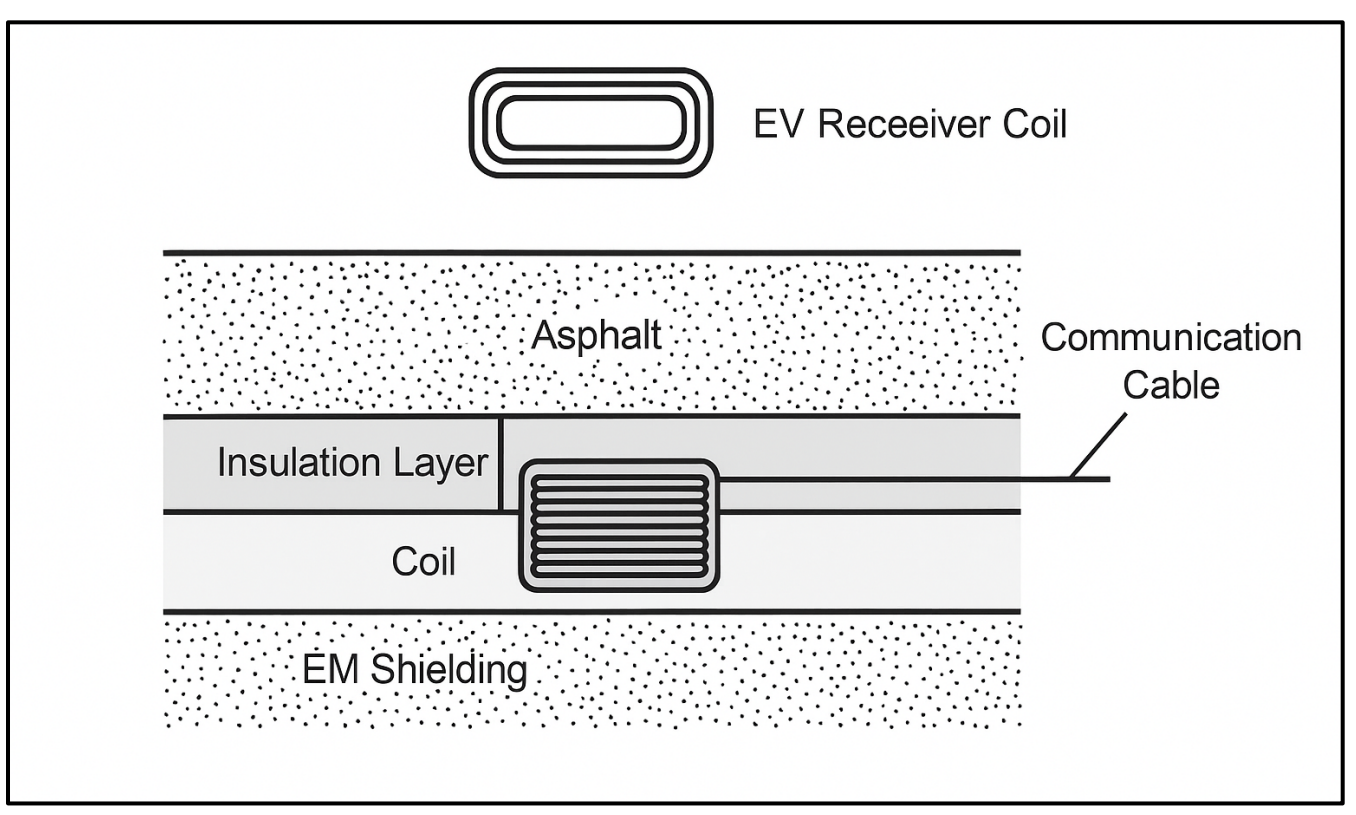

**Figure 1: Installation and Coil Structure – Cross-Section View**

### 3.2 Communication Layer

The communication layer uses low-latency V2I protocols to support real-time data sharing among EVs and RSUs. Data packets carry vehicle ID, battery SoC, requested energy, and billing dynamic metadata. 5G or future 6G technology provides stable, high-throughput communication [10]. Blockchain smart contracts can be used to support secure billing operations and privacy-preserving identity verification.
It is also responsible for executing traffic analytics, anomaly detection (e.g., energy theft) and predictive maintenance alerts. The RSUs send the summarised data to the cloud for advanced analytics and load forecasting.

### 3.3 Control & Grid Integration Layer

ERS operations are centrally managed using a cloud-based platform across various segments. Machine learning algorithms and artificial intelligence are used to forecast energy demand, manage load on the grid, and provide the highest priority to renewable sources (e.g., highway solar canopies). DERs and V2G are integrated to enable bidirectional energy transfer [11].

It consists of:

- Grid synchronisation using PMUs (Phasor Measurement Units)
- Load shedding during peak demand through control signals
- Renewable curtailment reduction through battery buffering systems

## 4 Methodology & Simulation

### 4.1 Simulation Setup

We developed a high-density urban corridor (5 km) with heterogeneous vehicle types by simulating with SUMO (Simulation of Urban Mobility) for traffic simulation and MATLAB for electromagnetic and power transfer analysis. The simulation was identified with traffic in an Indian city during peak hours (500–800 vehicles/hour).

**Table 1: Parameters and Values for Simulation**

| Parameter | Value |
|---|---|
| Coil spacing | 10 m |
| Coil efficiency | 88–91% |
| Activation time | < 200 m/s |
| Vehicle speed range | 20–70 km/h |
| Energy transfer per 2 km | 1.8–2.6 kWh |
| Peak coil load | 100 kW |

Key variables:

- Coil segment power: 50 kW
- Coil efficiency: 88–91%
- Speed range of vehicle: 30–60 km/h
- Battery SoC limit: 30–90%

We also modelled weather conditions, ambient temperature and seasonally varying road friction coefficients to estimate the effect of seasons.

### 4.2 Metrics Evaluated

**Energy delivered per segment (kWh):** Varied from an average of 1.6–2.8 kWh based on speed and SoC.
**Battery life:** Simulated over 1000 charge cycles. DWC cut deep discharge cycles by ~40%, enhancing battery life.
**Installation and maintenance cost:** Estimated at $1.8 million/km with 10-year lifecycle cost optimisation.
**Energy latency:** Minimal latency (<200 m/s) in coil activation/deactivation, ensuring smooth charging.
**Grid stability:** Deviation in voltage and frequency was within ±1.5%.

### 4.3 Mathematical Modelling of Power Transfer

The power transferred between the transmitting and receiving coils is governed by inductive coupling:

$$P = \eta \times Pt \qquad (1)$$

where:
P = received power
Pt = transmitted power
η = efficiency of power transfer

Efficiency depends on the coupling coefficient (k), mutual inductance (M), and the alignment:

$$\eta \propto k^2 \times Q \qquad (2)$$

where Q represents the quality factor of the coils.
The coupling coefficient varies with:
- Distance between coils
- Vehicle speed
- Lateral misalignment

### 4.4 Simulation Workflow

The simulation framework integrates traffic modelling and power analysis:

1. Traffic generation using SUMO
2. Vehicle speed and density extraction
3. MATLAB-based inductive power transfer modelling
4. Energy delivery computation per vehicle
5. AI-based optimisation using LSTM for load prediction

## 5 Results and Analysis

### 5.1 Power Transfer Behaviour

As shown in Figure 1, energy transfer efficiency is enhanced with medium velocities through increased coil coupling. A transfer rate of 2.6 kWh for 2 km was achieved at 50 km/h, which is the efficiency of the system during urban-speed performance.

**Fig. 2: Charging Efficiency vs. Vehicle Speed**

Charging efficiency peaks between 45 and 60 km/h due to optimal coil coupling. At lower speeds, energy transfer is limited

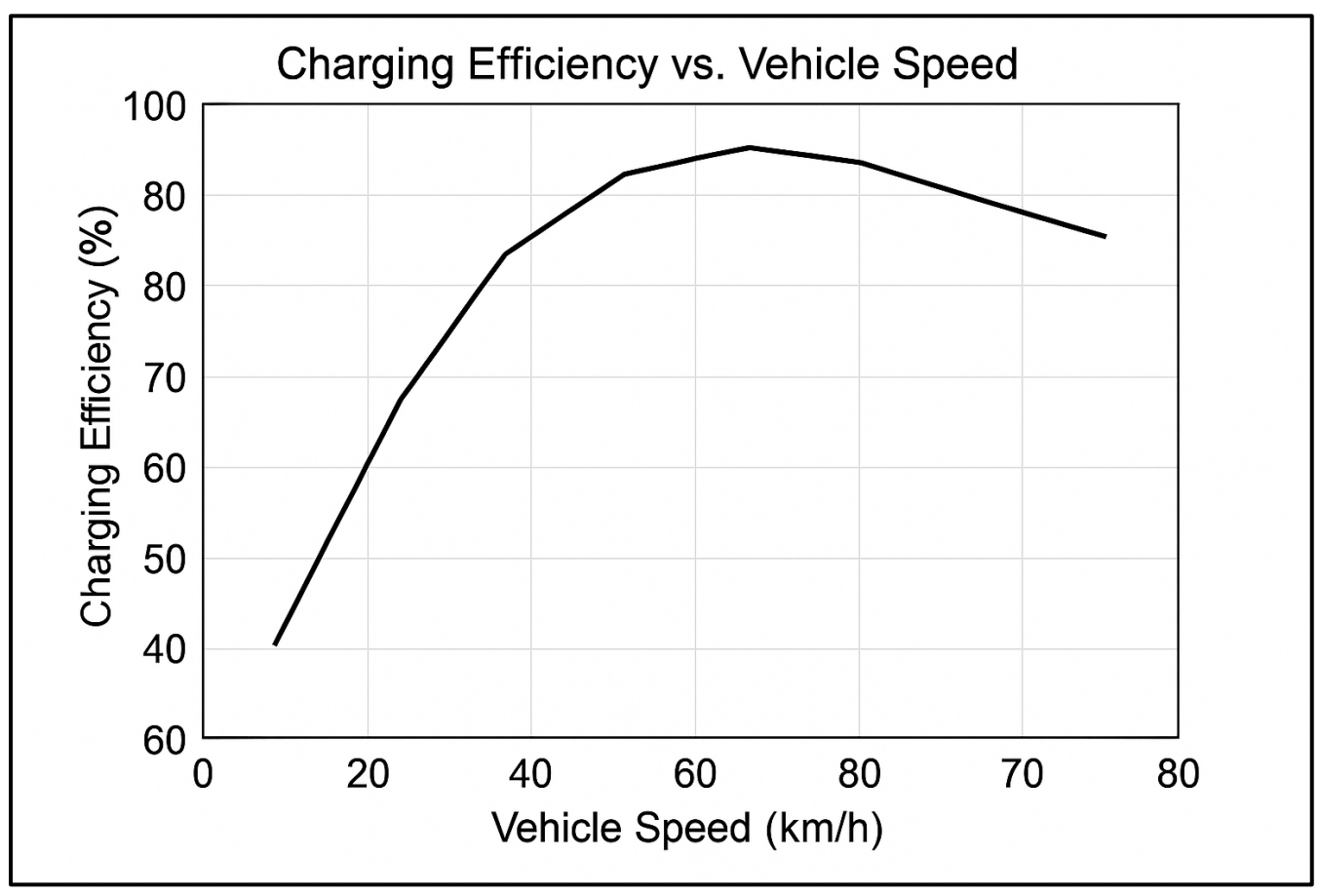

by shorter activation cycles, while at higher speeds, misalignment reduces efficiency.

### 5.2 System Architecture Overview

As shown in Figure 2, the layered structure of the system is Real-time charging is performed by the physical layer, security and metadata by the communication layer, and synchronisation with the smart grid by the control layer.

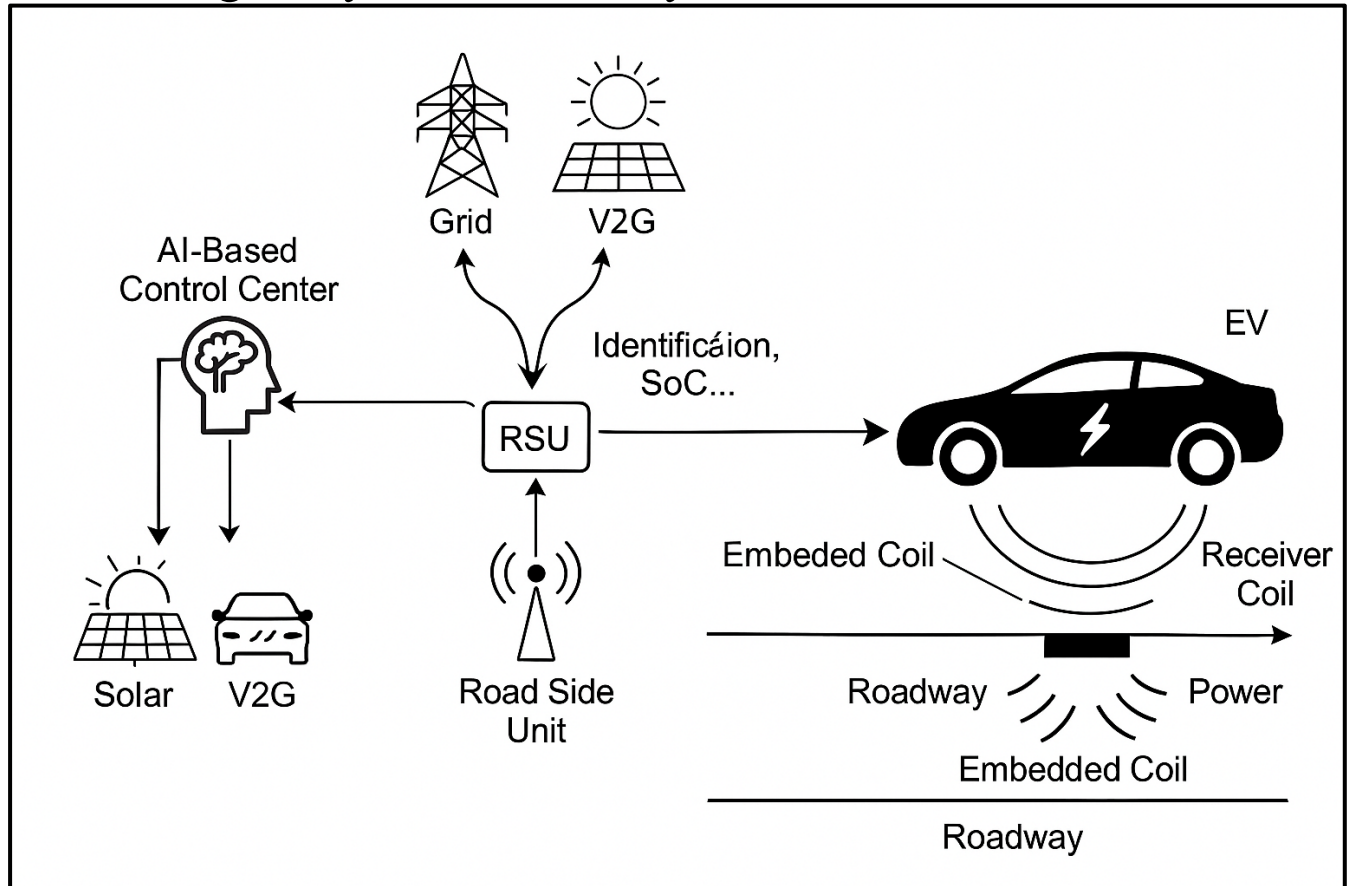

**Figure 3: ERS Power Transfer and System Control Architecture**

### 5.3 Financial Evaluation

While the cost of capital is still high (~$1.8 million/km), our financial model indicates break-even in 6–8 years, particularly for public transport and delivery fleets running on fixed city routes. Integration with India's FAME-II and green credits could bring this further down.

### 5.4 Battery Degradation Reduction

**Table 2: Effects on Battery (Static Vs ERS)**

| Metric | Static Fast Charging | ERS Dynamic Charging |
|---|---|---|
| Avg. Battery Temp (°C) | 45–50 | 32–36 |
| Deep Discharge Cycles | 310/year | 125/year |
| Estimated Battery Life | ~6 years | ~9 years |

### 5.5 Grid Load Optimisation and AI Control

Using an AI Energy Management System (EMS), this system distributes power to all Electrified Roadway System (ERS) segments in real time. The EMS uses a Long Short-Term Memory (LSTM) neural network to perform time series forecasting to predict future energy demand.

The LSTM model receives multiple data inputs, including:

- Historical traffic density (vehicles/hour)
- Time of day variations
- Solar irradiance levels
- Vehicle State of Charge (SoC)

The LSTM's output gives a prediction of how much energy each road segment will require in a short time period. This provides the EMS with the ability to:

- Dynamically activate/deactivate inductive coils
- Select the optimal source of energy (solar, grid, or V2G)
- Significantly reduce the peak power load at the grid

The LSTM model consists of three layers: input, hidden, and output. The LSTM was trained on simulated traffic-energy datasets created from vehicle simulation software (SUMO) and numerical simulation software (MATLAB) integrated together.

Results from the prediction model include:

- 94.6% accuracy in predicting hourly traffic flow
- Response latency of less than 150 m/s from the time a vehicle sends a signal to the time the coil is activated
- Utilisation of power sources as follows: 61% solar, 35% grid (off-peak), and 4% V2G input.

## 6 Case Study Delhi, India – Feasibility and Implementation Framework

Delhi, the capital of India, is known for congested traffic, poor air, and great electric vehicle (EV) policies. Outer Ring Road is more than 18 km long and interconnects major transport corridors, transit centres, and residential zones, with over 120,000 vehicles travelling along it every day. The road is constantly filled with cars, mainly buses, taxis, and delivery trucks, making it a critical location for frequent use of wireless charging.

### 6.1 Solar and Energy Potential

Delhi receives plenty of sunlight, around 2200 kWh/m² per year. This offers an excellent opportunity to utilise solar panels to supply ERS infrastructure on the corridors, parking lots, or dedicated charging stations. Grid-connected inverters could store excess energy during the day when sunlight is available to utilise during night hours or when sunlight is scarce.

By integrating solar power systems with smart inverters and battery storage systems:

- 60–70% of the energy requirements of ERS can be achieved by harnessing solar energy.
- Grid tension is reduced, and the cost of energy per km can be lowered below ₹1.7/km.

### 6.2 Pilot Deployment Target Zones

**Table 3: Pilot Deployment**

| Zone | Characteristics | Reason for Selection |
|---|---|---|
| IGI Airport–Mahipalpur stretch | High taxi & EV cab activity | 24/7 vehicle throughput, proximity to metro |
| Sarai Kale Khan ISBT | Dense public transport exchange | Frequent electric buses & auto-rickshaws |
| Ring Road Logistics Belt | Movement of goods carriers | Predictable fixed routes suitable for ERS |
| Nehru Place–Lajpat Nagar | Commercial & delivery hub | High-speed corridor with steady vehicular flow |

### 6.3 Environmental Impact

Positioning ERS at this site can provide these benefits:

- Yearly energy savings: ~9.8 GWh (for 18,000 EVs using ERS daily for 10 km)
- $CO_2$ reduction: Over 33,000 tons annually by consuming less petrol and diesel.
- Noise reduction: Continuous electric consumption lowers mean sound levels by 3–5 dB in urban areas.

### 6.4 Stakeholder Alignment

Delhi already has a friendly environment with:

- FAME-II incentives for electric fleets and buses.
- BSES and Tata Power Delhi as robust DISCOM partners.
- Transport Department, GNCTD electric vehicle policy.
- Pilot projects in the Gati Shakti Masterplan, National Smart City Mission, and Delhi EV Policy 2020.

### 6.5 Scalability Potential

If it works:

- Phase II can comprise NH-48, Dwarka Expressway, and Delhi–Meerut Expressway.
- Partnering with luxury ride-sharing operators (like BluSmart and Uber Green).
- Revenue-sharing contracts with fleet operators to recoup the installation costs.

## 7 Deployment Strategy: Phased and Scalable Rollout for ERS

Due to the capital intensity and urban sensitivities of ERS systems, phased introduction guarantees technological resilience, stakeholder acceptance, and minimum public disruption. Below is a three-phase deployment plan for Indian smart cities.

### 7.1 Phase I: Initial Test Deployment (1–2 km)

**Objective:** Proof-of-concept in a controlled, high-utilisation environment
**Location selection:** Sarais (as Sarai Rohilla, Sarai Kale Khan)
**Stakeholders:** Delhi city administration, Delhi Transco, Tata Power-DDL, and electric vehicle manufacturers.
**Parts:**

- 1 km single lane embedded coils
- 50 kW capacity solar microgrid
- Real-time edge computing module and RSU.

**Objectives:** Test wireless transfer speed (>90%), measure the interaction of grid load and traffic influence, and collect V2I communication performance data.
**Estimated Cost:** ₹15–20 crore (~$2M)
**Duration:** 12 months.

### 7.2 Phase II: Urban Corridor Integration (5–10 km)

**Objective:** Scale-up along high-frequency corridors with known traffic patterns
**Route type:** Outer Ring Road, Metro feeder roads, last-mile routes for e-buses
**Tech enhancements:**

- AI-based EMS for real-time coil activation and energy distribution.
- Smart contracts for fleet-specific billing via blockchain.

**Fleet Integration:**

- BluSmart, ETO Motors, DTC e-buses, Zomato EV delivery vans

**Metrics Tracked:**

- Total kWh transferred per vehicle type
- Time-in-coil (TIC) vs. energy received
- Traffic flow changes and driver experience feedback

**Support Mechanisms:**

- Involve NHAI and UTTIPEC for urban road re-engineering
- Green bond financing via smart city authorities
- Community outreach for awareness and safety

### 7.3 Phase III: Full Smart Grid-Integrated ERS

**Objective:** Create a self-sustaining, AI-optimised dynamic charging grid with V2G capabilities
**Coil coverage:** >25 km, covering the entire Ring Road and NH48 express stretch
**V2G Integration:**

- Public e-buses push energy into the grid during idle periods
- Time-of-day pricing and renewable-first grid logic

**National Integration**:

- Tie-in with National Smart Grid Mission (NSGM)
- Interoperability compliance with BIS and MoRTH standards

**Expected Outcome:**

- Full-scale reduction of 20–25% in urban EV charging infrastructure load
- Elimination of queue-based charging delays for fleets
- Replicable blueprint for Mumbai, Bengaluru, Ahmedabad, and Hyderabad

## 8 Conclusion

The paper has proved the implementation of a scalable architecture to provide highways with electricity to allow dynamic and wireless charging within a city. It provided simulation results to show a marked improvement in battery life and stability to the grid, and therefore an increase in efficiency of the way charging is performed.

The proposed system provides a means to increase range anxiety by up to 35% and will work with a distributed system for the management of energy, leveraging AI to optimise the energy use. A case study of Delhi demonstrated the economic feasibility of the proposed system and how scalable the system could be. Future work will include the development and deployment of an operational pilot for inductive charging.

## Acknowledgements

The authors would like to express their gratitude to academic mentors, transportation experts, and open-source communities for their valuable insights and tools that supported this research. We also acknowledge the use of simulation platforms such as SUMO and MATLAB, which enabled in-depth modelling of urban mobility and energy transfer. Special thanks to the Delhi Transport Corporation (DTC) for providing publicly available EV usage data that informed aspects of the case study. This work was conducted independently and was not supported by any formal funding or grant.